\documentclass[conference]{IEEEtran}
\IEEEoverridecommandlockouts

\usepackage{cite}

\usepackage{amsmath,amssymb,amsfonts}
\usepackage{algorithmic}
\usepackage{graphicx}
\usepackage{textcomp}
\usepackage{xcolor}
\usepackage{booktabs}
\usepackage{subcaption}
\usepackage{enumitem}
\usepackage{hyperref}
\def\BibTeX{{\rm B\kern-.05em{\sc i\kern-.025em b}\kern-.08em
    T\kern-.1667em\lower.7ex\hbox{E}\kern-.125emX}}
\begin{document}

\title{Overcoming Orchestration Bottlenecks at Exascale: A Decentralized, Policy-Driven Approach for Sim-AI Ensembles
 \thanks{This research used resources of the Argonne Leadership Computing Facility, a U.S. Department of Energy (DOE) Office of Science user facility at Argonne National Laboratory and is based on research supported by the U.S. DOE Office of Science-Advanced Scientific Computing Research Program, under Contract No. DE-AC02-06CH11357.}
}

 \author{\IEEEauthorblockN{Harikrishna Tummalapalli}
 \IEEEauthorblockA{\textit{Argonne Leadership Computing Facility} \\
 \textit{Argonne National Laboratory}\\
 Lemont, IL, U.S.A \\
 htummalapalli@anl.gov}
 \and
 \IEEEauthorblockN{Christine M. Simpson}
 \IEEEauthorblockA{\textit{Argonne Leadership Computing Facility} \\
 \textit{Argonne National Laboratory}\\
 Lemont, IL, U.S.A \\
 csimpson@anl.gov}
 \and
 \IEEEauthorblockN{Riccardo Balin}
 \IEEEauthorblockA{\textit{Argonne Leadership Computing Facility} \\
 \textit{Argonne National Laboratory}\\
 Lemont, IL, U.S.A \\
 rbalin@anl.gov}
 \and
 \IEEEauthorblockN{Vitali A. Morozov}
 \IEEEauthorblockA{\textit{Argonne Leadership Computing Facility} \\
 \textit{Argonne National Laboratory}\\
 Lemont, IL, U.S.A \\
 morozov@anl.gov}
 \and
 \IEEEauthorblockN{Thang D. Pham}
 \IEEEauthorblockA{\textit{Computational Science Division} \\
 \textit{Argonne National Laboratory}\\
 Lemont, IL, U.S.A \\
 tpham@anl.gov}
 \and
 \IEEEauthorblockN{Murat Keceli}
 \IEEEauthorblockA{\textit{Computational Science Division} \\
 \textit{Argonne National Laboratory}\\
 Lemont, IL, U.S.A \\
 kecelim@anl.gov}
 \and
 \IEEEauthorblockN{Thomas D. Uram}
 \IEEEauthorblockA{\textit{Argonne Leadership Computing Facility} \\
 \textit{Argonne National Laboratory}\\
 Lemont, IL, U.S.A \\
 turam@anl.gov}
 }

\maketitle

\begin{abstract}
Scientific computing is increasingly shifting from monolithic applications to coupled simulation-AI workflows composed of highly heterogeneous tasks with diverse hardware, scale, and runtime requirements. As these workflows scale to leadership-class systems, the resulting extreme ensemble sizes and task variability can create orchestration bottlenecks. System-level schedulers are often configured for limited throughput, while workflow tools face scalability issues due to rigid control-plane topologies and static scheduling heuristics. We introduce EnsembleLauncher, a recursively hierarchical workflow orchestrator for exascale systems, featuring a fully decentralized control plane and a programmable scheduling policy interface. On the Aurora supercomputer, EnsembleLauncher successfully scales to the entire machine with up to eight million serial tasks, outperforming state-of-the-art tools by more than four times. Additionally, we implement a programmable scheduling interface and demonstrate a significant impact of scheduling policies on resource utilization for high-variance ensembles and active learning pipelines representative of modern coupled simulation-AI workflows.
\end{abstract}


\section{Introduction}

AI-driven discovery is becoming a major part of scientific computing, moving workloads from monolithic, static simulations to dynamic, composite workflows that integrate simulation, inference, and learning \cite{bard2023workflow,brewer_ai-coupled_2025}. Large-scale MPI applications are increasingly being replaced by loosely coupled components, transforming high-performance computing (HPC) workloads into complex graphs of interdependent, heterogeneous tasks. These tasks span a large spectrum, ranging from short-running serial inference to long-running, multi-node MPI simulations. Recent examples of this shift include the active or online learning of ML surrogates \cite{dharuman_mprot-dpo_2024}, inverse design pipelines \cite{yan_mofa_2025, MLDocking}, digital twins \cite{brewer_digital_2024,henneking_real-time_2025}, and multi-agent workflows \cite{sinclair_scalable_2025, pham_chemgraph_2026}.

On leadership-class exascale systems, such as the Aurora \cite{aurora} and Frontier \cite{atchley_frontier_2023} supercomputers, the sheer scale of scientific exploration can result in millions of concurrent tasks distributed across thousands of nodes for these workflows, necessitating high-throughput scheduling. System-level schedulers, such as Slurm and PBSPro, are configured to orchestrate large allocations and can cause bottlenecks under these extreme throughput demands \cite{zhou_exploring_2013}. To bypass this, small jobs are often bundled into single, large allocations using batch scripts; however, this approach is rigid and extremely difficult to generalize for highly heterogeneous task ensembles.

User-space workflow systems have traditionally been designed to bridge this gap. These tools range from domain-specific programming languages \cite{di_tommaso_nextflow_2017,wozniak_swiftt_2013} to specialized orchestration frameworks \cite{zaharia_accelerating_2018,evans_exascale_2022}. Architecturally, they span a wide spectrum from centralized orchestrators that interface with the system scheduler \cite{salim_balsam_2019, deelman_evolution_2019} to fully decentralized frameworks \cite{moritz_ray_2018, wozniak_swiftt_2013}. While a centralized architecture simplifies state tracking and consistency, its centralized control loop can severely limit the task throughput required by modern AI-coupled workflows. Conversely, existing decentralized tools either lack native support for multi-node MPI applications \cite{moritz_ray_2018} or impose prohibitively steep learning curves on domain scientists.

Scheduling in parallel task systems is NP-hard \cite{du_complexity_1989}, yet existing workflow tools rarely expose a programmable scheduling interface, preventing domain scientists from injecting the custom runtime policies required to address the specific bottlenecks of their unique workflows.



This paper addresses both the throughput and scheduling issues by developing an exascale-capable tool targeting emerging AI-coupled workflows. With this work, we make the following contributions: 
\begin{enumerate}[leftmargin=*]
\item We develop EnsembleLauncher, a recursively hierarchical workflow orchestrator featuring a fully decentralized control plane and a programmable scheduling and task routing policy interface for exascale Simulation-AI (Sim-AI) ensembles.
\item Analyzing the performance of two independently developed frameworks, we demonstrate that control-plane topology rather than implementation is the primary determinant of task throughput at scale.
\item We evaluate EnsembleLauncher on the Aurora supercomputer, scaling to 8,192 nodes with 8 million serial tasks and outperforming state-of-the-art tools by more than four times.
\item Using EnsembleLauncher's programmable scheduling interface, we demonstrate the significant impact of scheduling policies on resource utilization for representative modern Sim-AI workflows.
\end{enumerate}


\section{Design Motivation}
\label{sec:design_motivation}

Modern AI-driven scientific campaigns are highly heterogeneous, interleaving vastly different computational resource requirements within a single workflow. For example, multi-agent frameworks like StructBioReasoner \cite{sinclair2025scalableagenticreasoningdesigning} and ChemGraph \cite{pham_chemgraph_2026} orchestrate external API calls for LLM reasoning alongside local single-GPU structure predictions and multi-core CPU analysis. Inverse design pipelines, such as MOFA \cite{yan_mofa_2025}, push this heterogeneity further by tightly coupling localized single-GPU inference tasks with multi-node atomistic simulations.

The temporal variance within these workflows is equally extreme. Task durations in MOFA range from sub-second generative inference tasks (0.1 to 0.5 seconds) to long-running, multi-node atomistic simulations lasting over 1,500 seconds. In contrast, the MLDocking \cite{vasan_high_2024} workflow exhibits an inverted temporal profile, where cumulative inference tasks require $\mathcal{O}(1000s)$ to screen billions of compounds while multi-node pre-processing tasks and docking simulations complete in $\mathcal{O}(10s)$. To support these disparate workloads on leadership-class supercomputers, an orchestrator must seamlessly co-schedule granular serial tasks and massive MPI allocations while sustaining high throughput and low dispatch latency.

To motivate the architectural choices of EnsembleLauncher, we conducted a controlled scaling study using homogeneous, serial tasks. Each task occupies a single CPU core and executes a fixed-duration busy-wait to simulate active computation. We configured each node for a concurrency of 64 and set the task count to ten times the concurrency. Task durations of 0.1s and 60s were selected to span the disparate workload extremes previously identified, while maintaining a bounded overall test duration. All evaluations were performed on the Aurora supercomputer at Argonne National Laboratory, where each compute node features two Intel Xeon Max Series CPUs providing a combined 104 physical cores.

We selected Dask as a representative orchestration framework due to its widespread adoption across HPC and data science pipelines. To isolate the impact of control-plane configuration on throughput, we evaluated Dask under two configurations:

\begin{enumerate}[leftmargin=*]
    \item Depth = 0: A flat configuration deploying one Dask worker per core, forcing all individual workers to communicate directly with the centralized scheduler.
    \item Depth = 1: A hierarchical topology deploying one Dask worker per compute node, with each worker managing a local pool of 64 processes. In this configuration, only a single aggregate entity per node communicates with the central scheduler.
\end{enumerate}

\begin{figure}
    \centering
    \includegraphics[width=\linewidth]{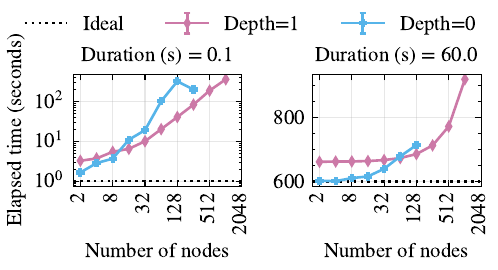}
    \caption{Elapsed time as a function of number of nodes for Dask tests with two choices of execution hierarchies.}
    \label{fig:scaling_motivation}
    \vspace{-12pt}
\end{figure}

Figure \ref{fig:scaling_motivation} illustrates the overall elapsed time as a function of the node count. At small scales ($\le$ 32 nodes), both configurations exhibit comparable performance. However, at larger node count, the flat configuration (Depth-0) degrades rapidly and fails to execute beyond 256 nodes. Conversely, the hierarchical configuration (Depth=1) performs significantly better, confirming that reducing the fan-out communication bottleneck at the central scheduler is an absolute requirement for exascale. Still, with the single-level hierarchy evaluated here, the throughput scalability wall is only shifted from $\mathcal{O}(10^2)$ to $\mathcal{O}(10^3)$ nodes, which is insufficient for exascale systems.

To identify the source of performance degradation, Table \ref{tab:latency_breakdown} decomposes the per-task elapsed time at 128 nodes into three components: scheduling overhead, i.e., the dispatch-to-execution delay corrected for expected local queuing; execution time; and completion detection latency. For a worker processing 10 sequential tasks, the ideal mean queuing delay is 4.5$\times$ the execution time; any excess represents orchestration overhead. For 0.1s tasks, the elapsed time with the flat topology is dominated by scheduling overhead and completion detection latency, leaving workers largely idle. The hierarchical configuration reduces this overhead by an order of magnitude by limiting central scheduler contention. For longer 60s tasks, scheduling overhead is still significant with the flat configuration, but becomes negligible with the hierarchical setup.

\begin{table}[h]
    \centering
    \begin{tabular}{cclcl}\toprule
         Task Duration (s)&  \multicolumn{2}{c}{Scheduling Overhead (s)}&  \multicolumn{2}{c}{Completion Detection (s)}\\
         &  Depth = 0&Depth = 1&
    Depth = 0&Depth = 1\\\midrule
 0.1& 27.6& 1.9& 25.19&2.9\\
 60& 14.05& 0.9& 0.9&0.4\\ \bottomrule\end{tabular}
    \caption{Task latency breakdown with Dask at 128 nodes.}
    \label{tab:latency_breakdown}
\end{table}

An orchestrator of Sim-AI ensembles must also deliver ultra-low dispatch latency, in addition to high throughput, to prevent worker starvation on sub-second inference tasks. State-of-the-art decentralized tools address this through bottom-up scheduling, resolving task placement locally before incurring network hops to remote nodes \cite{moritz_ray_2018}. A hierarchical architecture naturally accommodates this by allowing tasks to be injected directly at any level of the tree, including leaf nodes, bypassing the global routing logic entirely for latency-sensitive workloads.

Beyond throughput, the orchestrator must also determine the optimal execution schedule for highly heterogeneous task ensembles. This involves two fundamental decisions: task ordering, which determines the execution priority of queued tasks at each node, and task routing, which determines how tasks and resources are partitioned across the scheduling hierarchy. However, scheduling in parallel task systems is NP-hard \cite{du_complexity_1989}, making a universally optimal heuristic impossible. Since the effectiveness of any heuristic depends on the specific workload characteristics, the orchestrator must expose a programmable scheduling interface rather than hardcoding a single strategy.

Our analysis into modern scientific workflows and the scheduling overhead incurred with Dask establishes two key requirements for an exascale Sim-AI orchestrator: (1) a recursively hierarchical and configurable topology; (2) support for co-scheduling heterogeneous ensembles of serial inference and multi-node MPI tasks, and (3) a programmable scheduling interface for workload-dependent task ordering and routing policies. These requirements motivate the architectural design of EnsembleLauncher.

\section{System Design}

\begin{figure}[h!]
    \centering
    \includegraphics[width=\linewidth]{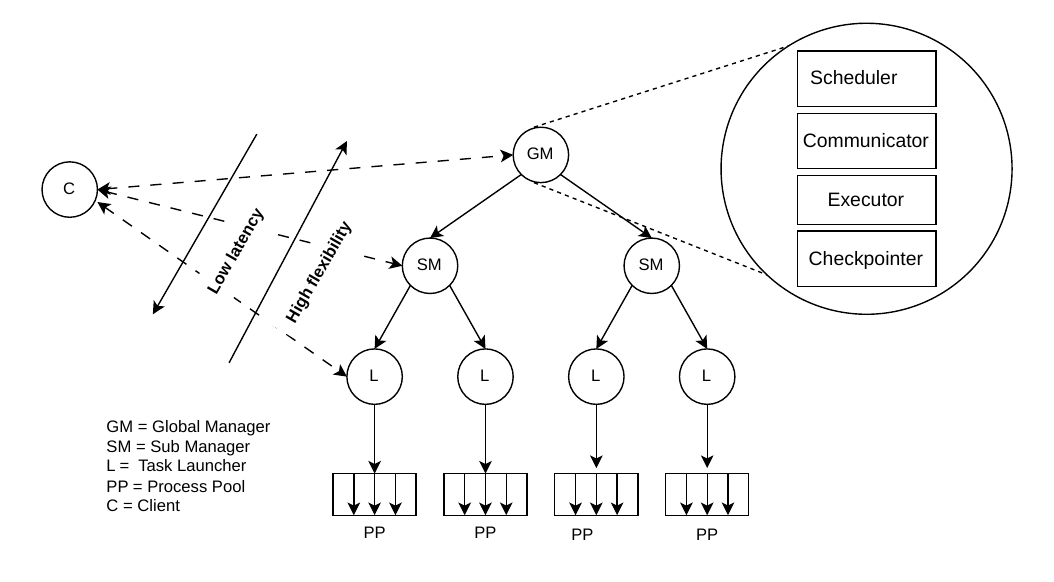}
    \caption{Architectural overview of EnsembleLauncher.}
    \vspace{-12pt}
    \label{fig:architecture}
\end{figure}

As illustrated in Figure \ref{fig:architecture}, the control plane is structured as a hierarchical tree composed of two orchestrator types: Managers, which govern subtrees, and Launchers, which execute tasks. The Global Manager serves as the root, intermediate Sub-Managers enable recursive nesting to arbitrary depth, and Launchers serve as leaf nodes.

Each orchestrator contains five core components: an executor, a scheduler, a resource manager, a communicator, and a checkpointer. The scheduler operates at two levels: coarse-grained task routing at the Manager level and fine-grained task ordering at the Launcher level. Both levels expose a programmable policy interface.

EnsembleLauncher can be deployed in two modes. In batch mode, the entire set of tasks are specified upfront and the system terminates upon completion. In cluster-client mode, the orchestrator hierarchy runs as a persistent service, accepting dynamic task injection to any orchestrator within the hierarchy via a client interface.

\subsection{Manager}

Managers govern the lifecycle and state synchronization of downstream orchestrators. The Global Manager serves as the root entry point, while intermediate Sub-Managers govern their respective children. Because these children can be either another layer of Sub-Managers or terminal Launchers, the design is inherently recursive, enabling arbitrarily deep scheduling topologies. Each Manager aggregates three subsystems: a Communicator for bidirectional messaging with its parent and children; a Children Scheduler to control the topology and task routing of the subtree; and an Executor to spawn child processes during expansion phases.

To accommodate diverse workload characteristics, the Manager operates in two distinct modes. In \textbf{Static Partitioning (Push)} mode, designed for batch processing, the task set is pre-partitioned during initialization and pushed to children only during setup, minimizing runtime control-plane communication overhead. In \textbf{Dynamic Load Balancing (Pull)} mode, designed to mitigate execution time skew, the manager employs a work-stealing protocol, responding to asynchronous task requests from idle children.

Uniquely, the system allows these modes to be mixed within the same hierarchy. By enforcing Static Partitioning at the upper levels and enabling Dynamic Load Balancing exclusively at the leaf tier (depth $d_{\text{max}}-1$), we establish a hybrid push-pull architecture that localizes high-frequency synchronization traffic to the leaf level, effectively isolating the root from communication bottlenecks. In pull mode, the Manager activates a per-child asynchronous monitor that triggers the local scheduling policy upon receiving a \texttt{TaskRequest}, ensuring requests are processed in parallel without blocking.

\subsection{Launcher}

The Launcher is the leaf orchestrator in the hierarchy, serving as the primary execution engine. Its core responsibility is to translate logical task definitions into physical execution via the Task Scheduler and the configured Executor.

To achieve low dispatch latency without burning idle CPU cycles, the Launcher implements a fully asynchronous, event-driven architecture for task execution. A dedicated submission monitor awaits state changes in the Task Scheduler's dispatch queue, as described in Section \ref{sec:task_scheduler}. If a schedulable task is promoted, the monitor wakes up and immediately offloads the execution to the backend Executor. Critically, resource lifecycle management is handled via non-blocking callbacks: upon task completion, a callback triggers the resource de-allocation routine. This instantly frees hardware resources for subsequent tasks without blocking the main event stream.

In pull mode, the Launcher complements this event-driven execution with periodic load balancing: a background monitor generates a \texttt{TaskRequest} to its parent Manager when the local queue falls below a depletion threshold, and an asynchronous listener ingests incoming \texttt{TaskUpdate} payloads to ensure newly assigned work is scheduled with low latency.

\subsection{Communicator}
\label{sec:communicator}

The Communicator component is engineered to deliver low-latency message exchange while maintaining strict backend extensibility. To support diverse deployment environments, it abstracts the underlying transport layer, seamlessly interfacing with protocols such as ZeroMQ \cite{hintjens2013zeromq} or standard multiprocessing pipes via a unified API. This modular design minimizes the integration burden, requiring custom backends to implement send/receive primitives for parent-child communication, and an optional heartbeat mechanism. By default, we enable the ZeroMQ transport backend with a heartbeat mechanism.

The ZeroMQ transport employs two distinct connection endpoints. The primary endpoint multiplexes diverse operational traffic, including control signals (e.g., Start, Stop) and task/result payloads. The secondary endpoint is dedicated exclusively to high-frequency heartbeat signals, isolating node health-checks from heavy data transfers. An asynchronous listener awaits incoming traffic on both endpoints. Upon detecting a message, the system instantly demultiplexes and dispatches the payload to an internal routing queue for downstream processing.

\subsection{Scheduler and Resource manager}

\subsubsection{Children Scheduler}
\label{sec:children_scheduler}

The Children Scheduler governs resource partitioning and task routing throughout the EnsembleLauncher hierarchy. This component utilizes a programmable policy engine to dynamically determine the configuration of child nodes based on the available resource pool and workload characteristics. Formally, the scheduling decision is a mapping:
\begin{equation}
    \label{eq:child_policy}
    S_{i+1} = f(S_i, T, N, d),
\end{equation}
where $S_i$ represents the state of the child pool (tasks and nodes assigned to each child) at allocation iteration $i$, $T$ denotes the set of pending tasks, $N$ represents the available compute nodes, and $d$ indicates the current hierarchy depth. The policy function $f$ is responsible for either allocating tasks to existing children (state mutation) or provisioning a new child (topology expansion) to accommodate the workload. 

To ensure extensibility, the architecture requires any valid custom policy to implement two core primitives: node partitioning and task assignment. We currently provide three reference implementations. Simple Splitting subdivides resources into a fixed number of children and distributes tasks among them in a round-robin fashion. Fixed Leaf Count Splitting ensures a fixed number of leaf nodes, computing all intermediate node counts via $\log_2$-space interpolation, coupled with round-robin task distribution. Hybrid Greedy Splitting greedily partitions resources to accommodate the most resource-intensive tasks first, followed by a round-robin distribution for the remaining lighter tasks.

Operationally, when EnsembleLauncher is executed in batch mode, the Children Scheduler policy is invoked exclusively at job initialization. In contrast, in cluster-client mode, the task routing logic is dynamically evaluated every time a new task arrives at the Manager or Launcher.

\subsubsection{Task Scheduler}
\label{sec:task_scheduler}
The Task Scheduler manages fine-grained execution at the leaf level of the hierarchy. To maximize resource utilization and minimize scheduling overhead, this component adopts a strictly event-driven architecture. An asynchronous monitor awaits resource availability signals; the moment hardware capacity is freed, the scheduler identifies eligible tasks that fit the current footprint and promotes them from the pending priority queue to the dispatch queue. The ordering of this pending set is governed by a programmable prioritization policy.

Formally, the task scheduling policy is a mapping:
\begin{equation}
    \label{eq:launch_policy}
    P = f(T, S),    
\end{equation}
where $f$ represents the policy function, $T$ denotes the task metadata, $S$ denotes the internal scheduler state, and $P$ is the resulting priority scalar used for queue ordering. We currently provide a reference implementation that prioritizes the most resource-intensive tasks to optimize localized bin-packing efficiency. Furthermore, the policy interface exposes a callback mechanism, triggered upon task completion. This dynamic feedback loop facilitates the active learning and real-time refinement of the scheduling heuristics.

\subsubsection{Resource Manager}

The Resource Manager is a component of the scheduler that is instantiated and operates identically on every orchestrator in the hierarchy. To handle the diverse hardware found in modern HPC systems, we designed the manager as a simple accounting layer. It treats physical hardware as generic resource units. This abstraction allows the scheduling logic to remain completely decoupled from the underlying system architecture. The specific resource units tracked by this layer are determined by a user-defined node configuration set at initialization.

\subsection{Executor}
To address the heterogeneity of modern HPC workloads, the executor component is designed as a stateless abstraction layer that strictly decouples orchestrator (Manager/Launcher) state from the execution mechanism. This clear separation allows each task to be bound to a suitable executor at runtime. Additionally, the Children Scheduler policy can dynamically assign distinct executor backends to different branches of the hierarchy based on specific task requirements. This design enables EnsembleLauncher to co-schedule serial and MPI tasks at the Launcher level.

To ensure broad extensibility, the architecture defines a minimal interface contract requiring only \texttt{submit} and \texttt{shutdown} primitives, allowing users to integrate custom execution environments with minimal effort. As reference implementations, the system currently provides backends for a Python process pool, Python thread pool, MPI process pool, and direct MPI execution. The process pool enables efficient execution of serial tasks within a single node, while the MPI pool extends this capability to span multiple nodes.

Our MPI pool implementation operates by launching a single MPI communicator via an \texttt{mpiexec} command spawned through \texttt{subprocess.Popen}. The first rank of this communicator is designated as a gateway rank, responsible for relaying work between the main Python process and the other MPI ranks. Communication between the gateway and the main process is handled through Unix domain sockets using ZeroMQ.

\subsection{Checkpointing and Restart}
The checkpointer is the primary component enabling fault tolerance in EnsembleLauncher. Each orchestrator in the hierarchy maintains its own checkpointer, which periodically writes the current state to the parallel filesystem to enable restart. To minimize I/O overhead, only Launchers checkpoint task results, and only the Global Manager checkpoints task definitions. During initialization, each node searches for an existing checkpoint file and, if one is found, merges the recovered state with its current state to resume execution.

Fault detection relies on the heartbeat mechanism described in Section \ref{sec:communicator}. When a Manager detects that a child has become unresponsive, it automatically initiates a localized teardown and restart of the failed subtree. Conversely, when an orchestrator detects that its parent is dead, it propagates a stop signal down its subtree, checkpoints its current state, and terminates gracefully.

\subsection{Client Interface}

In the cluster-client configuration, the control plane operates as a long-running service that accepts tasks dynamically through the Client interface. Upon task submission, the Client immediately returns a \texttt{Future} object, conforming to the standard Python \texttt{concurrent.futures} API. 

While this interface matches the semantics of many existing Python-based execution frameworks, the novelty of our architecture lies in its topological flexibility: the Client can connect to \textit{any} orchestrator within the hierarchy, whether it is an intermediate Manager or a leaf Launcher. Tasks submitted to the Global Manager benefit from maximum scheduling flexibility, as they can be dynamically routed to any available resource across the cluster. However, traversing multiple hops through the tree structure inherently incurs a routing latency penalty. To circumvent this overhead for strictly latency-sensitive workloads, the client can connect directly to a local Launcher, bypassing the global routing logic entirely to ensure immediate, localized dispatch.

\begin{figure}
    \centering
    \includegraphics[width=\linewidth]{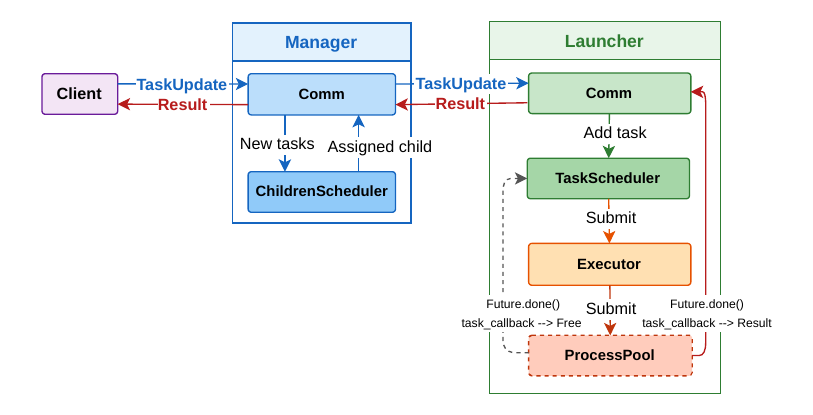}
    \caption{Dataflow diagram of Ensemble Launcher in cluster-client mode.}
    \vspace{-12pt}
    \label{fig:dataflow}
\end{figure}

To provide an overview of the task life cycle, Figure \ref{fig:dataflow} details the asynchronous execution pipeline within the cluster-client mode, illustrating step-by-step how task distribution and result resolution are routed through the system. While the diagram depicts a single-level hierarchy consisting of one Manager and one Launcher, this Manager routing flow is recursively repeated at every depth for deeper architectures. 

First, a \texttt{TaskUpdate} message from the Client is received by the Manager's Communicator. The newly arrived tasks are passed to the Children Scheduler, which assigns a specific child ID to route the update. This update is subsequently forwarded down the hierarchy and received by the Launcher's Communicator. Upon receipt, the Launcher adds the new tasks to the Task Scheduler's pending priority queue. 

Once hardware capacity is available, the Task Scheduler promotes the task and submits it to the Executor. Depending on the runtime configuration, the Executor routes the workload to a process pool, a thread pool, or a newly spawned subprocess. Upon task completion, a non-blocking callback mechanism is triggered by \texttt{Future.done()}. This callback instantly frees the local scheduling resources and transmits the result back through the Launcher's Communicator to the Manager. Finally, the Manager routes this result back up to the Client, successfully resolving the initially created \texttt{Future} object.

\section{Evaluation}
Here we evaluate the performance of EnsembleLauncher's on-node task throughput, as well as its weak and strong scaling. We particularly focus on the effect of control plane topology on task throughput.

All the tests were conducted on the Aurora supercomputer at the Argonne Leadership Computing Facility \cite{aurora}. Aurora is a 10,624-node HPE Cray-Ex based system consisting of 166 racks with 21,248 CPUs and 63,744 GPUs. Each node consists of 2 Intel Xeon CPU Max Series and 6 Intel Data Center GPU Max Series. Each Xeon CPU has 52 physical cores supporting 2 hardware threads per core, and each datacenter GPU is further divided into 2 tiles or stacks. Each node has 8 HPE Slingshot-11 NICs, and the system is connected in a Dragonfly topology.

\subsection{Performance}

\subsubsection{Single-node micro-benchmarks}
We first evaluate single-node task throughput to isolate the scheduling overhead of each framework. EnsembleLauncher is tested in two configurations: (1) a single Launcher managing all available physical cores via its internal process pool (EL(1)), and (2) a Manager with 102 Launchers, assigning one Launcher per core (EL(102)). For a fair comparison, we configured all other frameworks with 102 workers: Parsl using the \texttt{HighThroughputExecutor}, Dask using a \texttt{LocalCluster}, and Python's native \texttt{ProcessPoolExecutor} as an upper-bound baseline.

\begin{figure}[h]
    \centering
    \includegraphics[width=\linewidth]{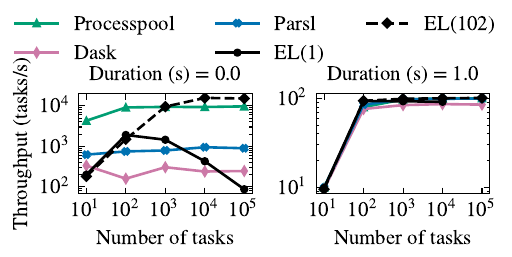}
    \caption{Single-node throughput comparison of Parsl, Dask, and EnsembleLauncher (EL) for varying task counts and durations.}
    \label{fig:worker_benchmark}
\end{figure}

Figure~\ref{fig:worker_benchmark} presents the single-node throughput as a function of task count for two task durations, 0s and 1s. At 0s duration, which isolates pure scheduling overhead, the Launcher configuration (one per node) achieves peak throughput at approximately 100 tasks (roughly one task per core) and sustains significantly higher throughput than both Parsl and Dask across the 100 to 1,000 task range. Beyond this point, however, throughput degrades. This is because the Task Scheduler computes a priority score and re-sorts the pending queue on every task completion, introducing overhead that grows with queue depth. This cost can be effectively mitigated by deploying multiple Launchers per node: the per-core Launcher configuration (EL(102)) demonstrates this, achieving throughput that approaches the native ProcessPool by distributing the scheduling overhead across independent Launcher instances. As task duration increases to 1s, the scheduling overhead is amortized and all frameworks converge toward the native ProcessPool baseline.

\begin{figure}[h]
    \centering
    \includegraphics[width=0.75\linewidth]{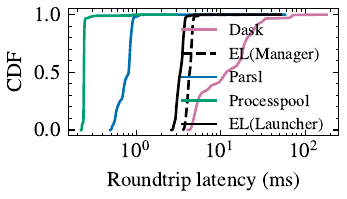}
    \caption{Cumulative distribution of task roundtrip latency.}
    \vspace{-15pt}
    \label{fig:latency_microbenchmark}
\end{figure}

We next measure task round-trip latency using the same single-node framework configurations as the throughput experiment, with one exception: EnsembleLauncher is tested with a single Manager and single Launcher rather than the 102-Launcher configuration in order to capture the latency introduced by one level of hierarchy. Round-trip latency is measured by submitting a single no-op task, recording the time from submission to result retrieval, and repeating this process for 10,000 tasks. Figure~\ref{fig:latency_microbenchmark} shows the cumulative distribution of the measured latencies. In the Launcher-only configuration, EnsembleLauncher achieves a median latency of approximately 3ms, substantially lower than Dask's 10ms. Adding one hierarchical level introduces roughly 1ms of additional latency, bringing the Manager-Launcher configuration to 4ms. Parsl achieves the lowest latency at 0.8ms, reflecting its optimized single-node executor path. Critically, all measured latencies remain well below the task durations typical of Sim-AI workflows, where even the shortest inference tasks run for tens to hundreds of milliseconds.

In summary, individual components of EnsembleLauncher are comparably efficient to existing frameworks in both throughput and latency. For typical workloads of approximately 1,000 tasks per Launcher, EnsembleLauncher competes with or outperforms existing tools and its overhead can be further amortized by deploying additional Launchers. Each hierarchical level adds approximately 1ms of latency, which remains well below the duration of even the shortest tasks observed in Sim-AI ensembles.

\subsubsection{Scaling}
\label{sec:performance_scaling}

We now evaluate the weak and strong scaling performance of EnsembleLauncher for an ensemble of serial tasks with uniform durations. In the weak scaling test, the ensemble size varies linearly with the number of nodes, and consists of 10 times the number of CPU cores. The strong scaling test fixes the total workload at 208,896 tasks and varies the node count from 64 to 2,048, corresponding to 3,264 down to 1 task per core. Unless otherwise noted, all reported measurements represent the mean of three independent runs; error bars denote the standard deviation.

For these tests, EnsembleLauncher is configured with a hierarchical depth of 2, ensuring exactly one Launcher per compute node. The distribution of children across the hierarchy is determined using the Fixed Leaf Count Splitting policy which performs $log_2$-space interpolation between the root node and the leaves. For instance, in an 8-node allocation, the hierarchy comprises 8 Launchers at depth 2, 2 Sub-Managers at depth 1 each connected to 4 Launchers, and the Global Manager at depth 0. We evaluate EnsembleLauncher in three configurations: batch mode with push scheduling (EL), cluster-client mode with push scheduling (EL(Cluster)), and cluster-client mode with pull scheduling (EL(Pull)). In all the EnsembleLauncher configurations, we use a process pool Executor with 102 workers at the Launcher level.

To ensure a fair comparison with state-of-the-art tools, we configured the baseline frameworks as follows. Parsl uses the HighThroughputExecutor with one Parsl manager and 102 workers per node (Depth=1). Dask is configured with one Dask worker per node utilizing 102 processes (Depth=1). The MPI baseline launches 102 processes per node, each identifying its assigned tasks via direct indexing; to account for the overhead of loading Python virtual environments at scale, it also executes a Python script. This configuration serves as the theoretical performance ceiling.

\begin{figure}[h] 
    \centering
    
    \begin{subfigure}[b]{\linewidth}
        \centering
        \includegraphics[width=\textwidth]{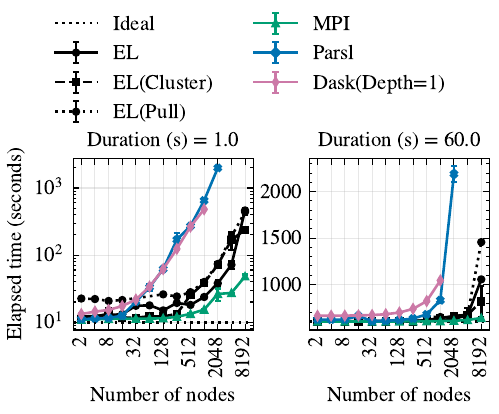}
        \caption{Weakscaling}
    \end{subfigure}
    
    \vspace{2pt} 

    \begin{subfigure}[b]{\linewidth}
        \centering
        \includegraphics[width=\textwidth]{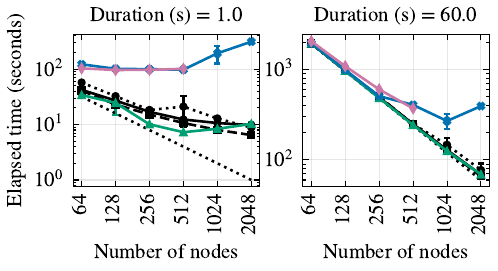}
        \caption{Strongscaling}
    \end{subfigure}
    
    \caption{Comparison of weak scaling (top) and strong scaling (bottom) performance for varying task durations.}
    \vspace{-15pt}
    \label{fig:scaling_overview}
\end{figure}

Figure~\ref{fig:scaling_overview} presents the weak scaling results up to 8,192 nodes on Aurora. EnsembleLauncher scales substantially better than both competing frameworks across all task durations. All the deployment configurations of EnsembleLauncher perform equally well. For 1s tasks, EnsembleLauncher outperforms Parsl and Dask at node counts as low as 8, and its elapsed time remains nearly flat through 128 nodes. Beyond this point, Dask fails to execute past 1,024 nodes, and Parsl was not tested beyond 2,048 nodes. At 2,048 nodes, EnsembleLauncher outperforms both frameworks by an order of magnitude. As task duration increases, the scalability wall shifts to higher node counts and scheduling overhead is amortized. However, even for 60s tasks, EnsembleLauncher outperforms the other frameworks by approximately 4$\times$ at 2,048 nodes. At 8,192 nodes, increased run-to-run variance is observed, primarily attributable to the exponential back-off mechanism implemented to avoid thundering herd effects during hierarchical teardown, where a parent orchestrator may take a variable duration to acknowledge child completion signals. Notably, this variance remains substantially smaller than the overall performance improvement over competing frameworks. A particularly noteworthy finding emerges from the direct comparison of Parsl and Dask. Despite being independently developed frameworks with fundamentally different internal architectures, they exhibit nearly identical scaling behavior when configured with the same single-level hierarchy, indicating the importance of the control plane topology.

The strong scaling results reinforce these findings. For 1s tasks at 2,048 nodes, EnsembleLauncher's scheduling overhead is an order of magnitude lower than that of the competing frameworks. For longer-duration tasks, EnsembleLauncher achieves near-ideal strong scaling performance.

In summary, EnsembleLauncher's recursive hierarchy delivers substantial scaling advantages over single-level frameworks, outperforming them by more than 4$\times$ at 2,048 nodes even for long-duration tasks and achieving near-ideal strong scaling.

\subsection{Topology sensitivity}
\label{sec:ablation}

\begin{figure}[h]
    \centering
    \includegraphics[width=0.85\linewidth]{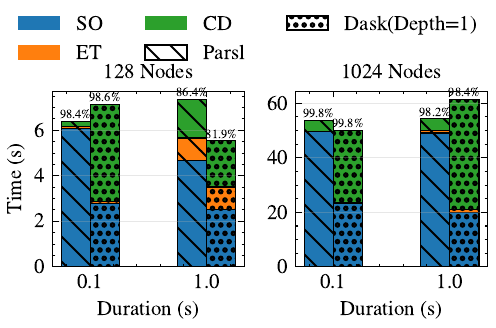}
    \caption{Mean Roundtrip latency breakdown comparison between Dask and Parsl.}
    \label{fig:latency_breakdown_ablation}
\end{figure}

In this section, we analyze the sensitivity of task throughput to control plane topology. Figure~\ref{fig:latency_breakdown_ablation} presents the per-task latency breakdown for Dask and Parsl at 128 and 1024 nodes, using the same three-component decomposition defined in Section~\ref{sec:design_motivation}: Scheduling Overhead (SO), Execution Time (ET), and Completion Detection (CD). The percentage of total time from orchestration overhead (scheduling overhead + completion detection) is annotated above each bar. These metrics are computed for each task; at 128 nodes, this equals 130,560 tasks.

Despite their fundamentally different architectures, both frameworks exhibit strikingly similar total orchestration overhead when configured with the same single-level hierarchy. The key difference lies in where the overhead manifests: Dask incurs substantially higher completion detection latency, while Parsl's overhead is concentrated in the scheduling phase. 

To isolate the effect of control-plane topology within a single framework, we configured EnsembleLauncher in two hierarchical configurations: a single-level hierarchy (Depth=1) mirroring the topology of Dask and Parsl, and a two-level hierarchy (Depth=2) introducing an intermediate Sub-Manager layer. Both configurations use the dynamic pull mode, in which idle Launchers issue explicit task requests to their parent. We also tested the effect of task-request batch size, comparing dynamically decided task request size against a constant size of 1,020 tasks.

\begin{figure}[h]
    \centering
    \includegraphics[width=0.75\linewidth]{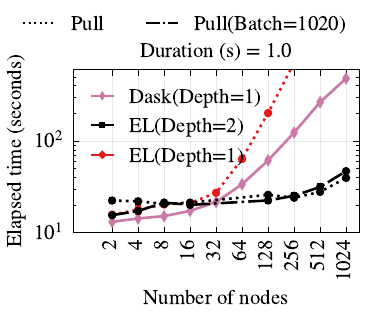}
    \caption{Performance comparison of EnsembleLauncher in various hierarchical configurations and Dask in a single level hierarchy.}
    \label{fig:scaling_ablation}
\end{figure}

Figure \ref{fig:scaling_ablation} shows scaling for 1.0s tasks. The single-level EnsembleLauncher configuration exhibits scaling behavior similar to Dask at the same depth, reinforcing our earlier finding that topology, not framework implementation, governs throughput at scale. The two-level configuration diverges from the single-level configurations at 32 nodes, maintaining near-ideal scaling through 256 nodes. Increasing the task-request batch size from 1 to 1,020 yielded no measurable improvement, suggesting that the additional communication overhead due to period task request and response pattern is effectively amortized by the hierarchy.

\begin{figure}[h]
    \centering
    \includegraphics[width=0.75\linewidth]{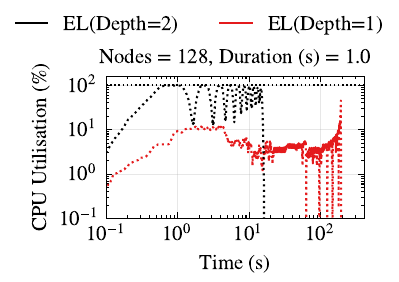}
    \caption{Comparison of resource utilization of various configurations of EnsembleLauncher.}
    \vspace{-12pt}
    \label{fig:resource_utilisation_ablation}
\end{figure}

Figure \ref{fig:resource_utilisation_ablation} illustrates the resource utilization over time for the same configurations at 128 nodes with 1.0s tasks. The two-level hierarchy (Depth=2) rapidly saturates all available cores and sustains near-full utilization throughout execution, completing in approximately 15 seconds. In contrast, the single-level configuration never fully saturates the available resources, never exceeding 10$\%$ of the available cores, resulting in substantial idle resources and an overall elapsed time approximately 5$\times$ longer than the two-level hierarchy. This demonstrates that the elapsed time degradation observed in Figure \ref{fig:scaling_ablation} is a direct consequence of control-plane saturation starving workers of schedulable work.

In summary, the topology sensitivity analysis confirms that control-plane topology is the primary determinant of task throughput at scale. Two independently developed frameworks converge to nearly identical overhead when configured at the same hierarchical depth, and adding a single additional level to EnsembleLauncher's hierarchy eliminates the control-plane saturation that starves workers at scale.

\subsection{Flexibility}
\label{sec:flaxibility}
In this section, we use EnsembleLauncher's Child Scheduler and Task Scheduler policies (Eqs. \ref{eq:child_policy} and~\ref{eq:launch_policy}) to demonstrate that scheduling policy can have significant impact on resource utilization and is workload-dependent.

\subsubsection{Heterogeneous bag-of-tasks}
To demonstrate the impact of task ordering we use a heterogeneous bag-of-tasks. This pattern is common in scientific campaigns, e.g. in parametric sweeps of design spaces to identify optimal configurations \cite{bobbitt_high-throughput_2016}, to construct datasets for AI/ML training \cite{zhao_core_2025,rosen_machine_2021}, and AI model hyper-parameter optimization \cite{deephyper}. Minimizing the makespan for such workloads is typically approached through heuristic-based scheduling; however, no single heuristic is universally optimal. We demonstrate this using a heterogeneous bag of tasks in which the node requirement of each task is sampled from a uniform distribution over 1 to 4 nodes, and the task duration is drawn from a Gamma distribution with a mean of 30s and a coefficient of variation ranging from 0.1 to 2.0 (corresponding to variances of 9$s^2$ to 3600$s^2$). The Gamma distribution is chosen for its strictly non-negative support, which naturally models computational task durations. To obtain statistical convergence each experiment is repeated 15 times. We evaluate four task scheduling policies, each governing the priority ordering at the Launcher level in Eq.~\ref{eq:launch_policy} and implemented as external modules plugged into EnsembleLauncher at runtime: \textbf{FIFO}, which uses strict first-in, first-out ordering; \textbf{Shortest First}, which prioritizes tasks with shorter estimated durations; \textbf{Longest First}, which prioritizes tasks with longer estimated durations; and \textbf{Largest First}, which prioritizes tasks with larger resource requirements.

\begin{figure}[h]
    \centering
    \includegraphics[width=0.75\linewidth]{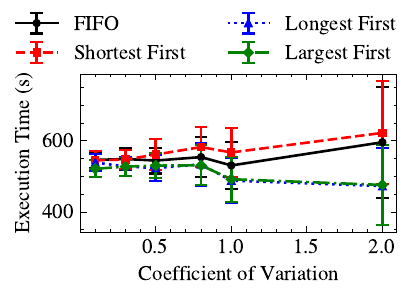}
    \caption{Execution time as a function a coefficient of variation for various heuristic based scheduling policies.}
    \label{fig:duration_vs_cv}
\end{figure}

Figure \ref{fig:duration_vs_cv} presents the total execution time of the ensemble as a function of the coefficient of variation (ratio of standard deviation to mean). At low variance, all policies yield comparable performance, as the task duration is approximately homogeneous. However, as the variance increases, the Largest First and Longest First policies achieve significantly lower run duration, demonstrating that scheduling policy choice can have a substantial impact on resource utilization in high-variance Sim-AI ensembles.

\subsubsection{Representative Sim-AI workflow}
\label{sec:pipeline}
To demonstrate the effect of resource partitioning and task routing policies on Sim-AI workflows, we developed a prototype mini-app of the MOFA workflow \cite{yan_mofa_2025}. The mini-app preserves the exact task duration ratios, fanout ratios, and resource requirements of the production pipeline. As in the real workflow, each stage of the pipeline is driven by an independent submitter process with dedicated input and output queues. To maintain a bounded test duration, all task durations are uniformly scaled down by a factor of 100, see Table \ref{tab:mofa}. While MOFA employs multiple tools within a single pipeline stage, the mini-app consolidates these into a single task whose duration equals the sum of the individual tool durations. The pipeline is seeded with 120,000 inference tasks at the initial stage, which, after successive filtering through the fanout ratios, yields a single estimate adsorption task at the terminal stage.

\begin{table}[h]
\centering
\caption{MOFA mini-app characteristics.}
\label{tab:mofa}
\begin{tabular}{llll}
\toprule
Task& Resources& Duration (s) & Fanout \\ \midrule
Generate linkers& 1 GPU &  0.0037 & 1.0\\
Process linkers& 1 CPU &  0.0012 & 0.228\\
 Assemble MOFs& 1 CPU & 0.03 & 0.125\\
 Validate structure& 1 GPU&2.240 & 0.086\\
 Optimize cells& 2 nodes&15.1270 & 0.0035\\
 Estimate adsorption& 1 CPU&21.0 & 1.0\\
 \bottomrule
\end{tabular}
\end{table}

EnsembleLauncher is deployed in a two-level hierarchy, with each Launcher node configured to use both the MPI and MPI pool executors concurrently. All serial tasks are tagged for execution via the MPI pool executor, while multi-node tasks are dispatched through the MPI executor. We evaluate three Children Scheduler policies. The \textbf{Uniform} policy assigns 2 nodes to each Launcher and distributes tasks in a round-robin fashion at all levels of the hierarchy. The \textbf{Partitioned} policy splits the hierarchy at the first level into three specialized Sub-Managers for multi-node simulation tasks, CPU-only tasks, and serial GPU tasks. Resources are partitioned according to the relative compute requirements, in GPU-seconds. At the second level, Launchers are placed at granularities of 2, 1, and 1 compute nodes for the three Sub-Managers, respectively. The \textbf{Adaptive} policy uses the same resource partitioning as Uniform but dynamically routes tasks to the least-loaded child, measured by the aggregate resource requirement of its currently assigned tasks. Notably, most often workflow tools use partitioned approach.

\begin{figure}[h]
    \centering
    \includegraphics[width=\linewidth]{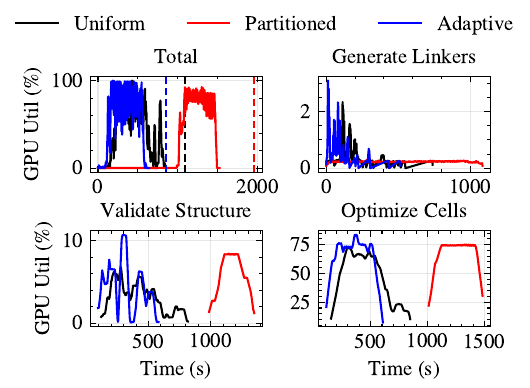}
    \caption{GPU utilization over time for the MOFA pipeline.}
    \label{fig:mofa_util}
\end{figure}

Figure \ref{fig:mofa_util} presents GPU utilization over time on 32 nodes, with vertical dashed lines marking the completion of each experiment. The Adaptive and Uniform policies achieve significantly higher utilization and shorter ($\text{2}\times$) durations than the Partitioned policy where the configuration funnels all serial inference tasks into a small dedicated subset of nodes. Since the initial inference tasks completely saturate the serial partition, the next set of serial GPU tasks (validate structure) wait in the queue, blocking the pipeline. The Uniform and Adaptive policies distribute inference tasks across the full node allocation, preventing this bottleneck. Of the two, the Adaptive policy achieves better performance by dynamically routing tasks to the least-loaded children, eliminating stragglers and maintaining balanced load across the hierarchy throughout execution.

In summary, both experiments demonstrate that scheduling policy choice is workload-dependent and can degrade resource utilization by up to 2$\times$ when mismatched to the workload characteristics. By exposing programmable interfaces at both the task ordering (Eq. \ref{eq:launch_policy}) and hierarchical routing (Eq. \ref{eq:child_policy}), EnsembleLauncher enables domain scientists to tailor scheduling behavior to their specific workflow without modifying the orchestrator.

\subsection{Limitations}
Our scaling experiments evaluate only homogeneous task ensembles. However, the flexibility study in Section~\ref{sec:flaxibility} demonstrates EnsembleLauncher's ability to optimize resource utilization for heterogeneous, multi-stage Sim-AI pipelines. These two evaluations are complementary: the scaling tests establish raw throughput at exascale, while the flexibility experiments validate the system's effectiveness under realistic workload heterogeneity.

EnsembleLauncher supports a hybrid push-pull mode, where static partitioning at the upper levels of the hierarchy and dynamic load balancing at the leaf level coexist. While this design localizes high-frequency synchronization traffic, it raises the concern of load imbalance. EnsembleLauncher's programmable routing policy directly addresses this by enabling dynamic task routing down the hierarchy. We demonstrated this capability through the Adaptive policy in Section \ref{sec:pipeline}. As future work, we aim to enable more aggressive cross-subtree load balancing through timeout-based mechanisms that detect straggler subtrees and resubmit pending tasks to less-loaded branches of the hierarchy.

The current fault tolerance mechanism is limited to checkpoint-restart at the subtree level: a failed orchestrator triggers teardown and restart of its entire subtree. Implementing fine-grained fault recovery that isolates failures to individual orchestrator remains future work.

\section{Related Work}

\subsection{System-Level Hierarchical Schedulers}

Several system-level schedulers have adopted partially hierarchical architectures \cite{hindman_mesos_2011,vavilapalli_apache_2013,foster_globus_1997,tannenbaum_condor_2001}, where a secondary sub-scheduler manages a static block of resources acquired from the global system manager. Flux \cite{ahn_flux_2018}, in contrast, introduces a fully recursive hierarchical resource management model. It allows schedulers to be nested arbitrarily deep within allocations, enabling a global instance to be recursively partitioned into multiple independent child schedulers. This design distributes scheduling overhead across the hierarchy, leading to significant improvements in maximum task throughput. However, Flux operates as a system-level resource manager where users submit jobs to a specific level of the Flux hierarchy, and the scheduler at that level allocates resources; there is no mechanism for routing tasks through the tree based on workload characteristics. EnsembleLauncher occupies a fundamentally different point in the design space. Its Children Scheduler policy (Eq. \ref{eq:child_policy}) enables the topology itself - tree depth, branching factor, and resource partitioning - to be programmatically determined as a function of the workload at initialization, and its routing policies govern how tasks are dynamically distributed through the hierarchy at runtime. As demonstrated in Section \ref{sec:pipeline}, the Adaptive routing policy achieved markedly higher GPU utilization than both Uniform and Partitioned configurations on the same allocation, underscoring the importance of workload-aware task routing, a capability that lies outside the scope of a system-level resource manager. 

\subsection{Application-Level Distributed Schedulers}

Ray was originally designed for reinforcement learning, and has emerged as a widely adopted industry standard for distributed Python execution \cite{moritz_ray_2018}. By utilizing a bottom-up distributed scheduling architecture, where local schedulers manage tasks at the node level and only forward metadata to a global control store when necessary, Ray achieves $\mathcal{O}(\text{ms})$ task latencies and high throughput. 
    
However, Ray is primarily optimized for elastic, loosely-coupled machine learning workloads (e.g., hyperparameter tuning, model serving). It faces significant challenges when deployed in traditional HPC environments. Specifically, Ray lacks native primitives for efficiently co-scheduling static MPI applications alongside dynamic serial tasks \cite{yuan_malleable_2025}. This often forces users to statically partition resources, negating the benefits of dynamic orchestration.

On the extreme end of the scalability spectrum, Swift/T utilizes a fully distributed, MPI-based architecture \cite{wozniak_swiftt_2013}. By compiling workflow definitions into scalable MPI programs via the Turbine runtime \cite{wozniak_turbine_2012}, it leverages the Asynchronous Dynamic Load Balancing (ADLB) library \cite{lusk2010more} to serve as a distributed work queue. This design eliminates centralized scheduling bottlenecks entirely, demonstrating peak throughput exceeding 1.5 billion tasks per second \cite{armstrong_compiler_2014}.

Despite this unparalleled performance, Swift/T faces significant challenges in co-scheduling traditional MPI applications alongside fine-grained serial tasks, a defining characteristic of modern AI workflows. This limitation stems primarily from the fundamental complexities of dynamically splitting and managing MPI communicators. Furthermore, Swift/T relies on a custom Domain Specific Language (DSL). This requirement imposes a steep learning curve on researchers and differentiates the workflow definition from the standard Python data science ecosystem, hindering widespread adoption in modern AI-driven research.

\section{Conclusion}
This paper presented EnsembleLauncher, a recursively hierarchical workflow system designed to overcome orchestration bottlenecks when executing heterogeneous Sim-AI ensembles on exascale systems. Through controlled scaling studies, we established that control-plane topology, not framework-specific implementation, is the primary determinant of task throughput at scale, with two independently developed frameworks exhibiting convergent scaling behavior at identical hierarchical depth.

EnsembleLauncher employs a fully decentralized, recursively hierarchical, control plane to eliminate the single-node serialization bottleneck that limits existing tools to $\mathcal{O}(10^3)$ nodes. On the Aurora supercomputer, EnsembleLauncher scaled to 8,192 nodes with up to 8 million serial tasks, outperforming state-of-the-art frameworks by more than 4$\times$. Note that 8,192 nodes is only the maximum allowable allocation on Aurora supercomputer, and not the scalability limit of EnsembleLauncher.

By leveraging EnsembleLauncher's programmable scheduling interface, we further demonstrated that scheduling policy choice has a significant impact on resource utilization and execution time for proxy Sim-AI coupled workflows. For high-variance task ensembles, Largest First and Longest First policies substantially outperformed FIFO as duration variance increased. In a multi-stage pipeline modeled after the MOFA workflow, adaptive, load-aware routing policies achieved markedly higher GPU utilization than both uniform and statically partitioned configurations, underscoring the importance of workload-specific scheduling logic for complex Sim-AI pipelines.

In future work, we plan to implement a Model Context Protocol (MCP) interface on top of the client interface, enabling LLM-driven agentic workflows to seamlessly execute tasks in leadership-class systems, thus bridging the gap between emerging AI agent frameworks and leadership-class HPC resources. Additionally, leveraging the programmable Task Scheduler and Children Scheduler policy interfaces, we plan to explore AI-driven scheduling and routing policies that learn from observed workload behavior and adapt their heuristics at runtime, potentially reducing makespan and improving resource utilization for highly dynamic Sim-AI campaigns.

\section*{Acknowledgment}

Generative AI models (Claude Opus 4.6 and Gemini Pro 3.5) have been used to improve the readability of the text.

\bibliographystyle{IEEEtran}
\bibliography{references}

\end{document}